\documentclass[%
 preprint,
 amsmath,amssymb,
 aps,
 pre,
showkeys
]{revtex4-2}
\usepackage{amsmath}
\usepackage[utf8]{inputenc}
\usepackage{graphicx}
\usepackage{dcolumn}
\usepackage{bm}
\usepackage{hyperref}
\usepackage{xcolor}
\usepackage{ctable}
\usepackage{url}
\usepackage{cancel}

\begin{document}

\title{Nonuniversal beyond-LHY corrections to thermodynamic properties of a weakly interacting Bose gas}
\author{Pham Duy Thanh}
\email[]{thanhdpham@outlook.com}
\author{ Nguyen Van Thu}
\affiliation{Department of Physics, Hanoi Pedagogical University 2, Hanoi, Vietnam}


\date{\today}

\begin{abstract}
We investigate the effects of finite-range interatomic interactions on the equation of state (EoS) of a weakly interacting Bose gas. Within the Cornwall–Jackiw–Tomboulis effective action approach, we show that finite-range effects influence not only the EoS but also the thermodynamic properties of the system at zero temperature, leading to nonuniversal behavior. A comparison with experimental data for the energy density shows an excellent agreement.
\end{abstract}


\maketitle

\section{Introduction\label{sec1}}

The study of thermodynamic properties of ultracold Bose gases dates back to the 1950s, initiated by the seminal works of Lee, Huang, and Yang (LHY) \cite{lee1957eigenvalues,lee1957many}. An important result of these studies is the LHY correction, which shows that the ground-state energy at zero temperature density acquires a contribution scaling as $(128/15\sqrt{\pi})\sqrt{\rho a_s^3}$, where $a_s$ is the $s$-wave scattering length and $\rho$ is the atomic density. Numerous subsequent studies have confirmed this finding \cite{brueckner1957bose,beliaev1958energy,lieb1963exact,haugset1998thermodynamics}. Recent years have witnessed a resurgence of interest in beyond–LHY correction in weakly interacting Bose systems, largely motivated by advances in the experimental realization of highly controllable quantum gas mixtures with tunable interactions \cite{mordini2020measurement,skov2021observation,busley2022compressibility,cominotti2023ferromagnetism}. Within the framework of the Cornwall–Jackiw–Tomboulis (CJT) effective action formalism at the two-loop level, corrections to the energy density up to order $(\rho a_s^3)^{3/2}$ have recently been reported \cite{van2022condensed}.

The CJT effective action approach developed in Refs. \cite{van2022condensed,zhang2024cornwall} has been shown to be a powerful and versatile framework for describing ultracold quantum gases. In particular, its application to Bose-condensed systems offers two major advantages: (i) it provides a set of self-consistent equations whose solutions are intimately connected to the mechanism of spontaneous symmetry breaking; and (ii) it enables a direct and systematic determination of thermodynamic quantities through the effective potential density. Importantly, Ref. \cite{zhang2024cornwall} demonstrates that this framework provides a natural and systematic approach to nonuniversal effects arising from finite-range interatomic interactions \cite{braaten2001nonuniversal,cappellaro2017thermal,tononi2018condensation,lorenzi2023shell,ye2024nonuniversal,yu2024interaction}.

As a continuation of previous studies employing the CJT effective action approach at two-loop level, in this work we investigate the nonuniversal beyond-LHY corrections to thermodynamic properties of a weakly interacting Bose gas. The paper is organized as follows: In Section \ref{sec 2}, after briefly reviewing the determination of the effective mass by using the CJT theory within improved Hartree–Fock (HF) approximation, we analyze several thermodynamic quantities in Section \ref{sec 3}. Finally, conclusions and perspectives for future work are presented in Section \ref{sec 4}.

\section{CJT effective potential density and effective mass}\label{sec 2}

Let us begin with the Lagrangian density of a Bose-condensed gas~\cite{zhang2024cornwall},
\begin{align}\label{001}
	\mathcal{L}[\psi,\psi^*]=&\ \psi(\textbf{r},t)\left(-i\hbar\frac{\partial}{\partial t}-\frac{\hbar^2}{2m}\nabla^2\right)\psi(\textbf{r},t)-\mu|\psi(\textbf{r},t)|^2+\frac{g}{2}|\psi(\textbf{r},t)|^4\nonumber\\
	&-\frac{\lambda}{2}|\psi(\textbf{r},t)|^2\nabla^2|\psi(\textbf{r},t)|^2,
\end{align}
where $\mu$ denotes the chemical potential, and $\psi(\mathbf{r},t)$ is the bosonic field depending on space $\mathbf{r}$ and time $t$; $m$ is the atomic mass. The coupling constant $g$ characterizes the interatomic interaction. 
Within the contact-interaction approximation, it is determined by the atomic mass $m$ and the $s$-wave scattering length $a_s$. 
Under the first Born approximation, the coupling constant is given by \cite{Pethick}
\begin{eqnarray}
g=\frac{4\pi\hbar^2 a_s}{m}.
\end{eqnarray}
The parameter $\lambda$ describes the finite-range correction and is defined as
\begin{eqnarray}
\lambda=\frac{2\pi\hbar^2 a_s^2 r_s}{m},
\qquad
r_s=a_s-\frac{r_e}{2},
\label{lambda}
\end{eqnarray}
where $r_e$ denotes the effective range of the interatomic interaction~\cite{fu2003beyond}, whose value depends on the specific form of the interaction potential. 
This quantity arises from the second-order term in the low-momentum expansion of the $s$-wave scattering amplitiude~\cite{Bethe1949}.

Then, we introduce the CJT effective potential density in the HF approximation that can be derived from Lagrangian density \eqref{001} in the manner that was pointed out in \cite{phat2009bose},
\begin{equation}\label{036}
	V_{\rm eff}[\psi_0,G]=-\mu\psi_0^2+\frac{g}{2}\psi_0^4+\frac{1}{2}\int_\beta\operatorname{Tr}[\ln G^{-1}(k)+G_0^{-1}(k)G(k)-1\!\!\!1]+\frac{3g}{8}\left(P_{11}^2+P_{22}^2\right)+\frac{g}{4}P_{11}P_{22}.
\end{equation}
Here, $\psi_0=\sqrt{\mu/g}$ denotes the expectation value of the field $\psi$ at the tree level. The quantities $P_{aa}$ are defined as the momentum integrals
\begin{equation}
	P_{aa}=\int_\beta G_{aa}(k)=\frac{1}{\beta}\sum_{n=-\infty}^\infty G_{aa}(\omega_n,k),\ a=(1,2),
\end{equation}
where $\beta=1/(k_BT)$ and $\omega_n=2\pi n\beta^{-1}$ are the bosonic Matsubara frequencies at temperature $T$ with $k_B$ the Boltzmann constant. The propagators $G_0(k)$ and $G(k)$ correspond to the tree-level and HF approximations, respectively, and are functions of the wave vector $\bf k$.

Subsequently, physical solutions are obtained by imposing the stationarity of the generalized effective potential, which is expressed through the conditions: $\delta V_{\rm eff}/\delta\psi(x)=0$, $\delta V_{\rm eff}/\delta G(x,y)=0$. It can be seen that central to the CJT theoretical framework is the treatment of the full propagator $G(x,y)$ as an independent variable, which allows one to derive self-consistent equations for both $\psi(x)$ and $G(x,y)$ from the CJT effective potential density. In particular, the variation $\delta V_{\rm eff}/\delta G(x,y)=0$ of the CJT effective potential $V_{\rm eff}$ in Eq.~\eqref{036} leads to a self-consistent Schwinger-Dyson equation of the form
\begin{equation}\label{003}
	G^{-1}(k)=G_0^{-1}(k)+\Sigma.
\end{equation}
On the right-hand side of Eq.~\eqref{003}, $G_0^{-1}(k)$ is the inverse propagator within one-loop approximation and take the form
\begin{equation}\label{029}
G_0^{-1}(k)=\begin{pmatrix}
	\varepsilon_k-\mu+3g\psi_0^2+2\lambda\psi_0^2k^2 & -\omega_n\\
	\omega_n & \varepsilon_k-\mu+g\psi_0^2
\end{pmatrix}
\end{equation}
where $\varepsilon_k=\hbar^2 k^2/(2m)$. Meanwhile, $\Sigma$ denotes the matrix containing the self-energies, namely
\begin{equation}
	\Sigma=\begin{pmatrix}
		\Sigma_1 & 0\\
		0 & \Sigma_2
	\end{pmatrix},
\end{equation}
with the self-energies $\Sigma_1$ and $\Sigma_2$ are given by
\begin{subequations}
	\begin{align}
		\Sigma_1&=\frac{3g}{2}P_{11}+\frac{g}{2}P_{22},\\
		\Sigma_2&=\frac{g}{2}P_{11}+\frac{3g}{2}P_{22}.
	\end{align}
\end{subequations}
It's easily to see that the propagator $G(k)$ can be obtain when turned out $G^{-1}(k)$, namely
\begin{equation}\label{004}
G(k)=\frac{1}{\omega_n^2+E^2(k)}\begin{pmatrix}
	\varepsilon_k-\mu+g\psi_0^2+\Sigma_2 & \omega_n\\
	-\omega_n & \varepsilon^{*}_k-\mu+3g\psi_0^2+\Sigma_1
\end{pmatrix}
\end{equation}
where $\varepsilon^*_k=\hbar^2 k^2/(2m^*)$, with modified mass $m^*$ is defined as
\begin{equation}
	m^*=\frac{m}{1+\dfrac{4m\lambda\psi_0^2}{\hbar^2}}\approx\frac{m}{1+8\pi\rho a_s^2 r_s}.
\end{equation}
In Eq.~\eqref{004}, $E(k)$ is the dispersion relation or the excitations spectrum, which is determined by examining poles of the inverse propagator,
\begin{equation}\label{035}
E(k)=\sqrt{(\varepsilon^{*}_k-\mu+3g\psi_0^2+\Sigma_1)(\varepsilon_k-\mu+g\psi_0^2+\Sigma_2)}.
\end{equation}

In the HF approximation, the excitations spectrum (\ref{035}) exhibits an energy gap, in violation of Goldstone’s theorem \cite{goldstone1962broken} in the context of spontaneous symmetry breaking. To remedy this issue and restore the Nambu-Goldstone boson, previous studies have supplemented the CJT effective potential density with a additional term, chosen such that the results in the symmetry-restored phase remain unchanged. As a consequence, the effective potential density takes the modified form
\begin{equation}\label{002}
	V_{\rm eff}[\psi_0,G]=-\mu\psi_0^2+\frac{g}{2}\psi_0^4+\frac{1}{2}\int_\beta\operatorname{Tr}[\ln G^{-1}(k)+G_0^{-1}(k)G(k)-1\!\!1]+\frac{g}{8}\left(P_{11}^2+P_{22}^2\right)+\frac{3g}{4}P_{11}P_{22}.
\end{equation}
Simultaneously, this modification results in an interchange of the self-energy definitions,
\begin{subequations}\label{007}
	\begin{align}
		\Sigma_1&=\frac{g}{2}P_{11}+\frac{3g}{2}P_{22},\\
		\Sigma_2&=\frac{3g}{2}P_{11}+\frac{g}{2}P_{22}.
	\end{align}
\end{subequations}
Analogous to the procedure used to determine the propagator in Eq.~\eqref{004}, the propagator within the improved HF approximation is given by
\begin{equation}\label{030}
	G(k)=\frac{1}{\omega_n^2+E^2(k)}\begin{pmatrix}
		\varepsilon_k & \omega_n\\
		-\omega_n & \varepsilon_k^*+M^2
	\end{pmatrix}.
\end{equation}
Clearly, the corresponding excitations spectrum is gapless and reads
\begin{equation}
	E(k)=\sqrt{\varepsilon_k\left(\varepsilon_k^{*}+M^2\right)}.
\end{equation}

By minimizing the CJT effective potential density in \eqref{002} with respect to the expectation value of the field operator, we derive the gap equation
\begin{equation}\label{008}
	-\mu+g\psi_0^2+\Sigma_2=0.
\end{equation}
In addition, the SD equation that the effective mass $M^2$ satisfies can be reformulated as
\begin{equation}\label{009}
	-\mu+3g\psi_0^2+\Sigma_1=M^2.
\end{equation}
It is important to note that, the chemical potential can be obtained by taking the first-order derivative of the $V_{\rm eff}$ in Eq.~\eqref{002} with respect to total density $\rho$,
\begin{equation}\label{013}
	\mu=-\frac{\partial V_{\rm eff}}{\partial\rho}=g\rho+gP_{11}.
\end{equation}
Combining Eqs.~\eqref{007} and \eqref{008}-\eqref{013}, we obtain
\begin{align}
	&\rho_{\rm ex}=\frac{1}{2}(P_{11}+P_{22}),\label{014}\\
	&M^2=2g\rho-3g\rho_{ex}-\frac{g}{2}(P_{11}-3P_{22}).\label{015}
\end{align}
Here, $\rho_{\rm ex}=\rho-\psi_0^2$ denotes the non-condensed density, also referred to as the quantum depletion.

To proceed, we must determine the self-energies; in other words, this amounts to evaluating the momentum integrals 
\begin{subequations}
	\begin{align}
		&P_{11}=\frac{1}{\beta}\sum_{n=-\infty}^\infty\int\frac{d\textbf{k}}{(2\pi)^3}\frac{\varepsilon_k}{\omega_n^2+E^2(k)},\\
		&P_{22}=\frac{1}{\beta}\sum_{n=-\infty}^\infty\int\frac{d\textbf{k}}{(2\pi)^3}\frac{\varepsilon_k^*+M^2}{\omega_n^2+E^2(k)}.
	\end{align}
\end{subequations}
By using the following formula
\begin{equation}
	\sum_{n=-\infty}^\infty\frac{1}{\omega_n^2+E^2(k)}=\frac{\beta}{2E(k)}\left(1+\frac{2}{e^{\beta E(k)}-1}\right),
\end{equation}
we have
\begin{subequations}\label{010}
	\begin{align}
		&P_{11}=\int\frac{d\textbf{k}}{(2\pi)^3}\frac{\varepsilon_k}{2E(k)}\left(1+\frac{2}{e^{\beta E(k)}-1}\right),\label{005}\\
		&P_{22}=\int\frac{d\textbf{k}}{(2\pi)^3}\frac{\varepsilon_k^*+M^2}{2E(k)}\left(1+\frac{2}{e^{\beta E(k)}-1}\right).\label{006}
	\end{align}
\end{subequations}
At zero temperature, the integrals \eqref{005} and \eqref{006} take the form
\begin{subequations}\label{021}
	\begin{align}
		&P_{11}=\frac{(2m^*)^{3/2}M^{3}}{6\pi^2\hbar^3}\sqrt{\frac{m^*}{m}},\label{021a}\\
		&P_{22}=-\frac{(2m^*)^{3/2}M^{3}}{12\pi^2\hbar^3}\sqrt{\frac{m}{m^*}}.\label{021b}
	\end{align}
\end{subequations}
To determine the parameter $M$, we first combine Eqs.~\eqref{014} and \eqref{015} and inserting the momentum integral $P_{11}$ given in in Eq.~\eqref{021a}. Finally, we readily find that $M$ obeys the following equation
\begin{equation}\label{018}
	M^2=2g\rho-\frac{(2m^*)^{3/2}g}{3\pi^2\hbar^3}\sqrt{\frac{m^*}{m}}M^3.
\end{equation}
We now briefly review the key steps involved in solving Eq.~\eqref{018} using perturbation theory. First, we introduce a small parameter $\epsilon$ such that
\begin{equation}\label{020}
	M^2=2g\rho-\epsilon\frac{(2m^*)^{3/2}g}{3\pi^2\hbar^3}\sqrt{\frac{m^*}{m}}M^3.
\end{equation}
Meanwhile, $M$ is expressed as a polynomial expansion in terms of the parameter $\epsilon$
\begin{equation}\label{019}
	M\to M_0 + \epsilon M_1 + \epsilon^2 M_2.
\end{equation}
Next, the expression for $M$ given in Eq.~\eqref{019} is substituted into Eq.~\eqref{020}, after which the resulting equation is organized in powers of $\epsilon$ up to $\mathcal{O}(\epsilon^2)$. This leads to
\begin{subequations}
	\begin{align}
		M_0=&\ \sqrt{2g\rho},\\
		M_1=&-\frac{(2m^*)^{3/2}g}{6\pi^2\hbar^3}\sqrt{\frac{m^*}{m}}M_0^2,\\
		M_2=&-\frac{M_1^2}{2M_0}-\frac{(2m^*)^{3/2}g}{2\pi^2\hbar^3}\sqrt{\frac{m^*}{m}}M_0 M_1.\label{028}
	\end{align}
\end{subequations}
Finally, we obtain the expression for $M$ as follows
\begin{align}\label{022}
	M=\sqrt{2g\rho}\left[1-\frac{16\sqrt{\rho a_s^3}}{3\sqrt{\pi}(1+8\pi\rho a_s^2 r_s)^2}\Bigg(1-\frac{40\sqrt{\rho a_s^3}}{3\sqrt{\pi}(1+8\pi\rho a_s^2 r_s)^2}\Bigg)\right].
\end{align}
The parameter $M$, as given in Eq.~\eqref{022}, plays a pivotal role in determining other thermodynamic quantities.

\section{Thermodynamics quantities}\label{sec 3}

\subsection{Quantum depletion and chemical potential}

Inserting $P_{11}$ and $P_{22}$ from Eqs.~\eqref{021} into Eq.~\eqref{014} yields
\begin{equation}\label{024}
	\rho_{\rm ex}=\frac{(2m^*)^{3/2}M^3}{24\pi^2\hbar^3}\left(2\sqrt{\frac{m^*}{m}}-\sqrt{\frac{m}{m^*}}\right).
\end{equation}
By substituting $M$ from Eq.~\eqref{022} into Eq.~\eqref{024}, we obtain the expression for the quantum depletion
\begin{align}\label{026}
	\rho_{\rm ex}=\rho\frac{8\sqrt{\gamma}}{3\sqrt{\pi}}\left(1-\frac{16\sqrt{\gamma}}{\sqrt{\pi}}+\frac{896\gamma}{3\pi}-\frac{24\pi r_s}{a_s}\gamma+\frac{640\sqrt{\pi}r_s}{a_s}\gamma^{3/2}\right).
\end{align}
where $\gamma=\rho a_s^3$ denote the gas parameter. The dilute-gas regime corresponds to $\gamma \ll 1$, meaning that the average interatomic distance $d\sim \rho^{-1/3}$ is much larger than the $s$-wave scattering length $a_s$, which characterizes the range of the interatomic interaction.

Next, inserting $P_{11}$ from Eq.~\eqref{021a} into Eq.~\eqref{013} yields
\begin{equation}\label{025}
	\mu=g\rho+\frac{(2m^*)^{3/2}M^3g}{6\pi^2\hbar^3}\sqrt{\frac{m^*}{m}}.
\end{equation}
By substituting $M$ from Eq.~\eqref{022} into \eqref{025}, the expression for chemical potential can be obtained systematically
\begin{equation}\label{027}
	\mu=g\rho\left(1+\frac{32\sqrt{\gamma}}{3\sqrt{\pi}}-\frac{512\gamma}{3\pi}+\frac{28672\gamma^{3/2}}{9\pi^{3/2}}-\frac{512\sqrt{\pi}r_s}{3a_s}\gamma^{3/2}+\frac{16384r_s}{3a_s}\gamma^2\right).
\end{equation}

When finite-range effects and finite-temperature contributions are neglected, our results coincide with those reported in Refs. \cite{van2022condensed,zhang2024cornwall}, except for the term of order $\gamma^{3/2}$ in the gas parameter expansion. This discrepancy can be readily traced to the omission of the term $M_2$ in $M$ [see Eq.~\eqref{028}] when substituting this parameter into the equations for the condensate depletion and the chemical potential, given by Eqs.~\eqref{026} and \eqref{027}, respectively.

To further analyze our results, we explore the nonuniversal quantum effects in the speed of sound, based on the chemical potential,
	\begin{equation}
		c=\ \sqrt{\frac{\rho}{m}\frac{\partial\mu}{\partial\rho}}=\sqrt{\frac{g\rho}{m}}\left(1+\frac{8\sqrt{\gamma}}{\sqrt{\pi}}-\frac{608\gamma}{3\pi}+\frac{50432\gamma^{3/2}}{9\pi^{3/2}}-\frac{640\sqrt{\pi}r_s}{3a_s}\gamma^{3/2}+\frac{29696r_s}{3a_s}\gamma^2\right).
	\end{equation}
\subsection{Pressure and ground-state energy density}
Another thermodynamic quantity of interest is the pressure, defined as the negative of the CJT effective potential \eqref{002} evaluated at its stationary point, where the gap and Schwinger-Dyson equations are simultaneously satisfied. Substituting Eqs.~\eqref{029} and \eqref{030} into Eq.~\eqref{002} yields
\begin{align}\label{031}
	\mathcal{P}=&\ \mu\psi_0^2-\frac{g}{2}\psi_0^4-\frac{1}{2}\int_\beta\operatorname{Tr}\left[\ln G^{-1}(k)\right]-\frac{1}{2}\left(-\mu+3g\psi_0^2-M^2\right)P_{11}-\frac{1}{2}\left(-\mu+g\psi_0^2\right)P_{22}\nonumber\\
	&-\frac{g}{8}\left(P_{11}^2+P_{22}^2\right)-\frac{3g}{4}P_{11}P_{22}
\end{align}
Combining \eqref{013}, \eqref{014} and \eqref{031}, the pressure expression \eqref{031} can be simplified to
\begin{equation}\label{032}
	\mathcal{P}=\frac{g\rho^2}{2}+g\rho P_{11}+\frac{g P_{11}^2}{2}-\frac{1}{2}\int_\beta \operatorname{Tr}\left[\ln G^{-1}(k)\right]
\end{equation}
To proceed further, the last term on the right-hand side of Eq.~\eqref{032} needs to be evaluated
\begin{align}\label{034}
	\frac{1}{2}\int_\beta \operatorname{Tr}\left[\ln G^{-1}(k)\right]=\frac{1}{2}\int\frac{d\textbf{k}}{(2\pi)^3}E(k)=\frac{(2m^*)^{3/2}M^5}{30\pi^2\hbar^3}\sqrt{\frac{m^*}{m}}.
\end{align}
Combining \eqref{032}, \eqref{034} and \eqref{021a} yields
\begin{align}\label{033}
	\mathcal{P}=\frac{g\rho^2}{2}\left(1+\frac{(2m^*)^{3/2}M^3}{3\pi^2\hbar^3\rho}\sqrt{\frac{m^*}{m}}-\frac{(2m^*)^{3/2}M^5}{15\pi^2\hbar^3 g \rho^2}\sqrt{\frac{m^*}{m}}+\frac{2m^{*3}M^6}{9\pi^4\hbar^6\rho^2}\sqrt{\frac{m^*}{m}}\right).
\end{align}
Substituting $M$ from Eq.~\eqref{022} into \eqref{033}, the expression for pressure can be derived
\begin{equation}\label{037}
	 \mathcal{P}=\frac{g\rho^2}{2}\left(1+\frac{64\sqrt{\gamma}}{5\sqrt{\pi}}-\frac{8192\gamma^{3/2}}{3\pi^{3/2}}-\frac{1024\sqrt{\pi}r_s}{5a_s}\gamma^{3/2}+\frac{4096r_s}{9a_s}\gamma^2\right).
\end{equation}


Another quantity of interest is the ground-state energy density, defined as the Legendre transform of the free-energy density $F(\mu)$, namely
\begin{equation}\label{039}
	\mathcal{E}=F(\mu)+\mu \rho = - \mathcal{P} + \mu \rho.
\end{equation}
Substituting Eqs.~\eqref{027} and \eqref{037} into \eqref{039}, we obtain the ground-state energy density
\begin{equation}\label{038}	\mathcal{E}=\frac{g\rho^2}{2}\left(1+\frac{128\sqrt{\gamma}}{15\sqrt{\pi}}-\frac{1024\gamma}{3\pi}+\frac{81920\gamma^{3/2}}{9\pi^{3/2}}-\frac{2048\sqrt{\pi}r_s}{15a_s}\gamma^{3/2}+\frac{94208r_s}{9a_s}\gamma^2\right).
\end{equation}

It is evident that Eqs.~\eqref{037} and \eqref{038} show the nonuniversal beyond-LHY corrections to the pressure and the ground-state energy density of a weakly interacting Bose gas, respectively, with finite-range effects taken into account. In particular, the terms of order $\gamma^{1/2}$ in Eqs.~\eqref{026}, \eqref{027}, \eqref{037}, and \eqref{038} correspond to the well-known LHY corrections to the quantum depletion, chemical potential, pressure, and ground-state energy density, respectively \cite{lee1957eigenvalues,lee1957many}. Experimentally, the LHY correction term in the ground-state energy density was measured by Navon \textit{et al.}, yielding a value of $4.5(7)$ \cite{navon2011dynamics}, which is in good agreement with the theoretical prediction $128/(15\sqrt{\pi})\approx 4.81$. In addition, upon truncating the expansion in the gas parameter $\gamma$, the nonuniversal terms agree exactly with the corresponding terms obtained in Refs.~\cite{cappellaro2017thermal,tononi2018condensation,sharma2022finite,zhang2024cornwall} using the functional path-integral method. This agreement indicates that our approach provides a significant improvement in accuracy. Furthermore, nonuniversal effects give rise to higher-order contributions in the gas parameter $\gamma$ to the ground-state energy density. As a result, our expression differs from that reported in Ref. \cite{van2022condensed}, highlighting the significance of our approach.

\begin{figure}[h]
	\centering
	\includegraphics[width=0.5\textwidth]{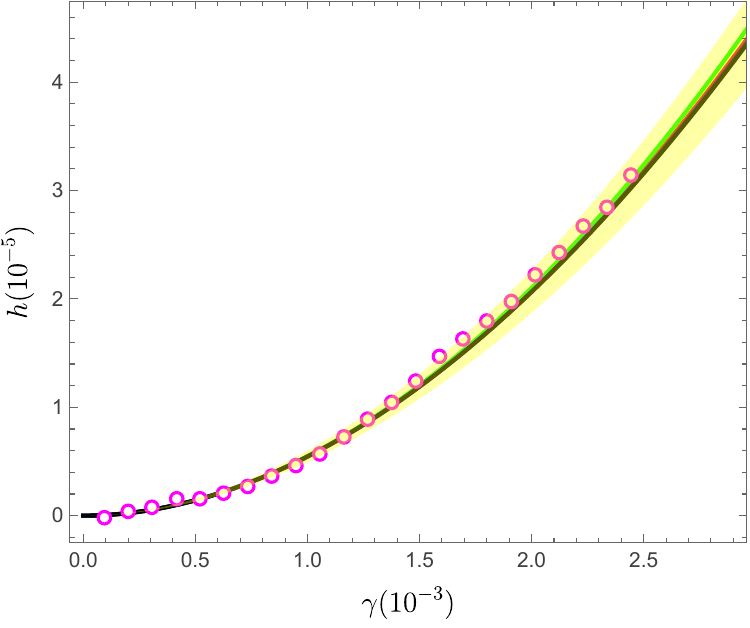}
	\caption{\label{fig1} The normalized pressure $h$ as a function of gas parameter $\gamma$ with different values of the dimensionless ratio $r_s/a_s$. The red and green lines correspond to $r_s/a_s=1/3, 1$, respectively. The black solid line shows the LHY correction. Here open circles are the experimental data in Ref.~\cite{navon2011dynamics} with a yellow error band.}
\end{figure}

To assess the experimental relevance of our results, we consider the case of $^{7}\mathrm{Li}$, as reported in Ref.~\cite{navon2011dynamics}. In this system, the scattering length $a_s$ and the gas parameter $\gamma$ are estimated to lie in the ranges $0 \sim 1.59 \times 10^{-7}\,\mathrm{m}$ and $0 \sim 3 \times 10^{-3}$, respectively. In addition, Ref.~\cite{wu2012optical} suggests that the effective range $r_s$ varies in the range $0 \sim \pm 3.71 \times 10^{-6}\,\mathrm{m}$, which allows us to explore the ratio $r_s/a_s$ within the range $0 \sim \pm 2$. To facilitate a direct comparison with experimental data, we introduce the normalized pressure
\begin{equation}\label{040}
h(\gamma) = \frac{\mathcal{P}}{\mu^2/2g} h_{\rm MF}(\gamma),
\end{equation}
where $h_{\rm MF}(\gamma) = 2\pi\gamma^2$ is the normalized pressure in the mean-field approximation. Fig.~\ref{fig1} shows the dependence of $h(\gamma)$ in Eq. (\ref{040}) on the gas parameter for several values of the ratio $r_s/a_s$. The experimental data from Ref.~\cite{navon2011dynamics} are represented by open magenta circles with a yellow error band. The black solid line corresponds to the standard LHY prediction, while the red line represents the case $r_e=2a_s/3$, equivalently $r_s/a_s=1/3$, associated with the hard-sphere interaction potential~\cite{lee1957many}. The green line corresponds to $r_e=0$, namely $r_s/a_s=1$, for which the effective-range contribution in the phase-shift expansion vanishes. For small values of the gas parameter, all theoretical curves are in good agreement with the experimental data, indicating that the system is well described within the weak-coupling regime. As the gas parameter increases, the deviations from the LHY prediction emerge, signaling the increasing importance of interaction effects beyond the standard contact-interaction approximation.
In particular, the finite-range correction associated with the hard-sphere potential leads to a visible improvement of the agreement with the experimental measurements in the strongly interacting regime. More interestingly, the case $r_e=0$ also provides a significant improvement over the pure LHY result. This observation demonstrates that the beyond-LHY correction is not determined solely by the effective-range term in the expansion of $k\cot\delta_0$. Even when $r_e=0$, the scattering amplitude still contains a nonvanishing momentum-dependent contribution proportional to $k^2$, originating from the intrinsic energy dependence of the two-body scattering amplitude. Consequently, the corresponding higher-order interaction correction remains finite and contributes appreciably at large gas parameters.

Finally, we consider the case of long-range van der Waals interactions between atoms~\cite{Gao1998}, characterized asymptotically by
\begin{eqnarray}
V(r)\sim -\frac{C_6}{r^6},
\qquad
r\rightarrow\infty,
\label{vander}
\end{eqnarray}
where $C_6$ denotes the van der Waals coefficient. 
For this class of interaction potentials, the effective range is given by~\cite{Gao1998}
\begin{eqnarray}
\frac{r_e}{\beta_6}
=
\frac{2}{3x_e}
\frac{1}{(a_s/\beta_6)^2}
\left\{
1+
\left[
1-x_e\left(\frac{a_s}{\beta_6}\right)
\right]^2
\right\},
\label{reC6}
\end{eqnarray}
where
\begin{eqnarray}
x_e=\frac{[\Gamma(1/4)]^2}{2\pi}\approx 2.0921,
\end{eqnarray}
and
\begin{eqnarray}
\beta_6=\left(\frac{mC_6}{\hbar^2}\right)^{1/4}
\end{eqnarray}
is the characteristic van der Waals length scale. In contrast to the hard-sphere model, Eq.~(\ref{reC6}) shows that the effective range is not an independent parameter, but depends nontrivially on the scattering length $a_s$. Consequently, in Eqs.~(\ref{037}) and (\ref{038}) for the pressure and energy densities, the ratio $r_s/a_s$ can be expressed in terms of the gas parameter through the combination $\tilde r_s/\gamma^{1/3}$, where
\begin{eqnarray}
\tilde r_s=\rho^{1/3}r_s.
\end{eqnarray}
For the ${}^{7}\mathrm{Li}$ experiment reported in Ref.~\cite{navon2011dynamics}, Eq.~(\ref{reC6}) yields an effective range of approximately $r_e=85a_0$, where $a_0$ is the Bohr radius. The corresponding dimensionless parameter is $\tilde r_s=0.136$. In this regime, the resulting curve for $h(\gamma)$ differs only marginally from the LHY prediction, indicating that the finite-range contribution associated with the van der Waals interaction remains weak for the experimentally accessible densities. 

\section{Conclusions}\label{sec 4}

In this paper, we have theoretically investigated nonuniversal quantum effects in an ultracold Bose-condensed gas with finite-range interatomic interactions within the IHF approximation based on the CJT effective field theory. Analytical expressions for several thermodynamic quantities have been derived as functions of the interaction parameters. Our results reveal the emergence of nonuniversal beyond-LHY corrections arising from quantum fluctuations and finite-range effects.

Importantly, we have shown that finite-range contributions can lead to measurable deviations in thermodynamic quantities, particularly in the ground-state energy density, thereby providing a feasible route for experimental detection of beyond-LHY effects.

Overall, our work offers a complementary perspective on ultracold quantum systems within the CJT formalism. The present approach can be further extended to investigate strongly correlated regimes and novel quantum phases, including quantum droplets in Bose mixtures as well as other ultracold Fermi systems.

\begin{acknowledgements}
We are very grateful to N. Navon for providing the experimental data.
\end{acknowledgements}

\section*{Conflict of interest}
All of the authors declare that we have no conflict of interest.

\end{document}